\documentclass[journal]{IEEEtran}

\usepackage{booktabs}
\usepackage{threeparttable}
\usepackage{amsmath,graphicx,cite,amssymb}
\usepackage{amsfonts,multirow,bm,array,setspace,stfloats}

\usepackage{graphicx,float,cite,amssymb}
\usepackage{graphics}
\DeclareGraphicsExtensions{.pdf,.jpeg,.png,.jpg}   
\usepackage{multirow,bm,bbm,array,setspace}
\usepackage{textcomp}
\usepackage{psfrag}
\usepackage{color}
\usepackage{pstricks,enumerate}
\usepackage{bbm}
\usepackage{subfigure}
\usepackage{xpatch}
\usepackage{algorithm,algpseudocode}
\makeatletter
\newcommand{\distas}[1]{\mathbin{\overset{#1}{\kern\z@\sim}}}%
\newsavebox{\mybox}\newsavebox{\mysim}
\newcommand{\distras}[1]{%
    \savebox{\mybox}{\hbox{\kern1pt$\scriptstyle#1$\kern1pt}}%
	\savebox{\mysim}{\hbox{$\sim$}}%
	\mathbin{\overset{#1}{\kern\z@\resizebox{\wd\mybox}{\ht\mysim}{$\sim$}}}%
}

\ExplSyntaxOn
\cs_new:Npn \bibColoredItems #1#2
  {
    \clist_map_inline:nn {#2} { \cs_new:cpn {bib@colored@##1} {#1} } 
  }
\ExplSyntaxOff

\newcommand\bib@setcolor[1]{%
  \ifcsname bib@colored@#1\endcsname
    \expandafter\color\expandafter{\csname bib@colored@#1\endcsname}
  \else
    \normalcolor
  \fi
}

\xpatchcmd\@bibitem
  {\item}
  {\bib@setcolor{#1}\item}
  {}{\PatchFailed}

\xpatchcmd\@lbibitem
  {\item}
  {\bib@setcolor{#2}\item}
  {}{\PatchFailed}

\makeatother
\ifCLASSINFOpdf
\else
\fi

\newtheorem{theorem}{Theorem}

\newtheorem{remark}{Remark}

\begin{document}
	%
\title{Enhancing Communications and Sensing Simultaneously by Zero-Order Optimization of MTS}
\author{
\IEEEauthorblockN{
    Wenhai Lai, \IEEEmembership{Member,~IEEE} and  Kaiming Shen, \IEEEmembership{Senior Member,~IEEE}
}
\thanks{
Paper accepted to IEEE Internet of Things Journal, June 2026. The work was supported by the NSFC under Grant 12426306. An earlier version of this paper was presented in part at IEEE
SPAWC 2025 [DOI: 10.1109/SPAWC66079.2025.11143403]. \emph{(Corresponding author: Kaiming Shen.)}

Wenhai Lai is with the School of General Education, Dalian University of Technology, China (email: wenhailai@dlut.edu.cn).
 
Kaiming Shen is with the School of Science and Engineering, The Chinese University of Hong Kong, Shenzhen, China (e-mail: shenkaiming@cuhk.edu.cn).
}
}

%


\maketitle

\begin{abstract}
Metasurface (MTS) comprises an array of meta-atoms, each reflecting and inducing a phase shift into the incident wireless signal. We seek the optimal combination of phase shifts across all the meta-atoms to maximize the channel strength from transmitter to receiver. Unlike many existing works that heavily rely on channel state information (CSI), this paper proposes a statistical approach to the phase shift optimization in the absence of CSI, namely blind configuration or zero-order optimization. The main idea is to extract the key features of the wireless environment from the received signal strength (RSS) data via \emph{conditional sample mean}, with provable performance. Furthermore, as a windfall profit, we show that the proposed blind configuration method has a nontrivial connection to phase retrieval which can be utilized for active sensing. In a nutshell, by configuring a pair of MTSs blindly without channel estimation, we not only enhance the channel strength to facilitate wireless communication, but also enable receiver to localize transmitter. All we need is the RSS data that can be readily measured at receiver. Our algorithm is verified in prototype systems in the 2.6 GHz spectral band. As shown in field tests, the proposed algorithm outperforms the benchmarks (e.g., MUSIC) in the active sensing task, and in the meanwhile raises the signal-to-noise ratio (SNR) significantly by about 10 dB.
\end{abstract}

\begin{IEEEkeywords}
Metasurface (MTS), phase shift, channel enhancement, active sensing, received signal strength (RSS), blind configuration, zero-order optimization, blackbox optimization.
\end{IEEEkeywords}

\section{Introduction}
\label{sec:intro}

\begin{figure}[t]
    \centering
    \includegraphics[width=0.7\linewidth]{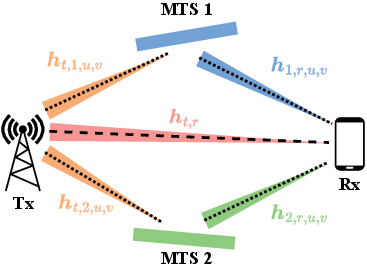}
    \caption{A pair of MTSs is deployed. We configure the phase shifts of the two MTSs to achieve the following two goals without CSI. First, enhance the wireless channel between transmitter and receiver. Second, enable receiver to localize transmitter based on the RSS, namely active sensing.}
    \label{fig:system_model}
\end{figure}

\IEEEPARstart{M}{etasurface} (MTS) \cite{al2017recent, Aobo2018MTS}, a.k.a. intelligent reflecting surface 
or reconfigurable intelligent surface,
is a planar array of meta-atoms. Each meta-atom is programmable in the sense that it induces a tunable phase shift into its reflected wireless signal. A typical application of MTS is to coordinate phase shifts across meta-atoms in order to focus the reflection beam onto the target receiver as shown in Fig.~\ref{fig:system_model}, thus boosting the signal-to-noise ratio (SNR). Most existing works assume that the phase shift optimization problem can be explicitly formulated as $\max\,f(x)$, so that the standard optimization methods based on the derivatives of $f(x)$ can be readily applied. Note that it entails the full channel state information (CSI) to write out this objective function $f(x)$.

\begin{figure*}
    \centering
\includegraphics[width=0.8\linewidth]{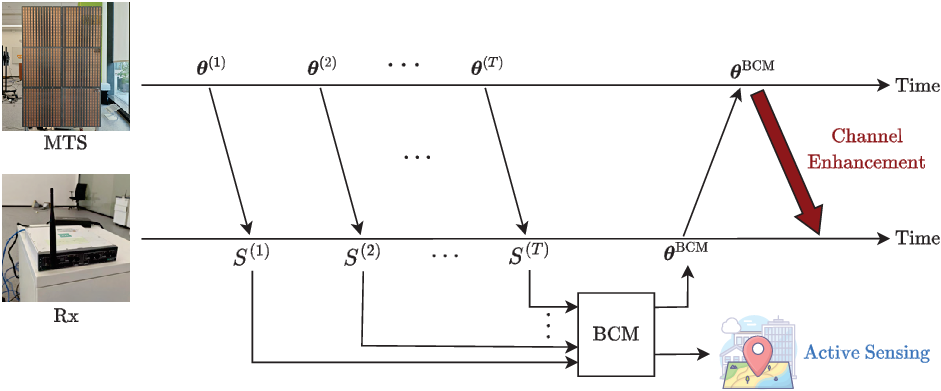}
    \caption{Procedure of proposed blind configuration method (BCM). Use a sequence of random phase-shift configurations $(\bm\theta^{(1)},\bm\theta^{(2)},\ldots,\bm\theta^{(T)})$ at the MTS side, and in the meanwhile measure the corresponding RSS values $(S^{(1)},S^{(2)},\ldots,S^{(T)})$ at the receiver side. The BCM will take the RSS dataset as input to produce two outputs: one is the position of the transmitter, and the other is the phase-shift solution $\bm\theta^{\text{BCM}}$ for MTS to enhance the propagation channel from the transmitter to the receiver. Note that the whole procedure does not involve the transmitter.}
    \label{fig:workflow}
\end{figure*}

In contrast, we propose a zero-order optimization method for MTS (which is called blind configuration) without any CSI knowledge, as illustrated in Fig.~\ref{fig:workflow}. This method however appears intractable at first glance; after all, we wish to optimize $x$ even without knowing the objective function $f(x)$. We show that blind configuration is statistically doable by extracting the key features of wireless environment from a set of random samples of the received signal strength (RSS). As another striking result, we establish a link between blind configuration and phase retrieval, based on which the receiver can localize the transmitter by treating a pair of MTSs as anchors. Thus, our blind configuration method can accomplish two tasks simultaneously: \emph{channel enhancement} and \emph{active sensing}. We would like to clarify that alongside the RSS data, the proposed algorithm strictly assumes prior knowledge of the receiver/MTS locations and requires controlled sampling of the MTS configurations to function properly.

With a phase shifter deployed at each meta-atom, one can manipulate the reflection beam of MTS to facilitate the wireless signal reception. A diverse range of application cases has been explored. For instance, \cite{scatterMIMO2020} uses MTS to improve multiple-input-multiple-output (MIMO) communication, \cite{cho2023mmwall} aims at the low earth orbit (LEO) satellite communication,  \cite{RFBouncer2023} considers using a dual-band MTS to enhance transmissions on two separate spectra simultaneously, \cite{chen2021pushing} uses MTS to address the polarization mismatch problem in wireless networks, \cite{chen2023seamless,Li2024RFMagus} use one or more MTSs to improve the wireless signal coverage,
\cite{Protego2022} suggests using MTS to generate artificial phase noise to prevent eavesdropping, and \cite{Feng2021RFlens} use MTS to increase the secrecy transmission rate. Moreover, a ``transparent'' MTS model is studied in \cite{cho2023mmwall,Arun2020RFocus}, by allowing each meta-atom to either reflect the incident signal or let it pass through. Many recent works have also investigated other types of MTS, such as absorptive MTS \cite{lin2024selfpowered}, refracting MTS \cite{an2024exploiting,lin2025aidriven,lin2022refracting}, and active MTS \cite{tang2026uncertainty}. This work considers the conventional MTS model as in \cite{xie2020max,li2024secure,Zargari2023energy,2020HuangcwJSAC, lai2024efficient,Xu2024Coordinating}; our main contributions lie in the dual-functional application (which aims to facilitate both wireless communications and active sensing) and the blind configuration method behind it.

Aside from the communication purposes, there is an emerging research interest in using MTS to improve localization. The authors of \cite{Li2024GPSwindow,2024GPMS} consider using the aforementioned transparent MTS to promote the indoor performance of GPS. The so-called MetaRadar scheme \cite{zhang2020metaradar} for indoor localization sets up MTS based on the RSS data. Our blind configuration method also takes the RSS data as input, but its rationale is quite different from the MetaRadar. The notion of ``anchor'' has played an important role in the indoor and outdoor localization \cite{gustafsson2005mobile,liu2007survey}. A natural idea is to utilize the MTS as an anchor. The authors of \cite{MetaSight2022} devise a so-called MetaSight system to localize a radio frequency identification (RFID) object by using a pair of MTSs. Its main idea is to retrieve the phase information for reflected channels and thereby recover the angle of arrival (AoA). With the AoAs from the different MTS, one can localize the object. It turns out that our blind configuration method (which is originally designed for the communication purpose) has an intimate connection to phase retrieval, so this method can provide a solution to active sensing as well.
In contrast, \cite{RIScan2024} suggests a beam scanning strategy for acquiring the AoA. Nevertheless, an analysis in this paper shows that beam scanning is less efficient than the proposed blind configuration method.

The phase shift optimization for MTS has been extensively studied in the literature to date. Many existing works assume that the full CSI is available and also that phase shifts can be arbitrarily chosen within the interval $[0,2\pi)$, so the phase shifting problem in their cases can be written explicitly as a continuous optimization problem. Two optimization tools have been quite popular because of the particular problem structure: the semidefinite relaxation (SDR) \cite{luo2010semidefinite,wu2019intelligent, zheng2021double, zhou2020robust, yao2023superimposed} and the fractional programming (FP) \cite{shen2018fractional_p1,feng2020physical,zhu2020power,zhang2022active,zhang2021joint}. However, in engineering practice, the choice of phase shift is typically limited to a discrete set due to the hardware cost, e.g., each meta-atom takes on its phase shift from the binary set $\{0,\pi\}$ in the 1-bit phase shifting case. To account for the discrete constraint, a common heuristic \cite{Feng2021RFlens,Protego2022,RFBouncer2023,wu2019beamforming,li2020joint,you2020channel} is to relax the problem and then round the continuous solution to the discrete set, but it lacks theoretical justification. As a somewhat surprising result from the recent work \cite{ren2022linear}, the phase shifts of a single MTS under the discrete constraint can be optimally determined within linear time; however, it is difficult to extend this global method for multiple MTSs. Unlike the above works, this paper advocates a blind configuration scheme that optimizes phase shifts in the absence of CSI. As discussed in \cite{Arun2020RFocus,ren2022configuring}, the following three facts motivate the MTS configuration without CSI in practice:
\begin{itemize}
    \item Each reflected channel is weak as compared to the background noise, so precise channel estimation is difficult.
    \item MTS consists of a large number of meta-atoms (e.g., our prototype has hundreds of meta-atoms), so channel estimation is computationally formidable.
    \item Channel estimation for MTS incurs extra overhead and requires modifying the current network protocol.
\end{itemize}
The main idea of the proposed blind configuration method (BCM) is to explore the wireless environment by analyzing the \emph{conditional sample mean} of the RSS data, which is a crucial statistical measure in the phase shift optimization. Although the use of conditional sample mean has been considered in the previous works \cite{Arun2020RFocus,ren2022configuring}, this work can be distinguished from the previous ones in two respects. First, we consider two or more MTSs, whereas \cite{Arun2020RFocus,ren2022configuring} are limited to a single MTS. Second, we consider a dual-functional objective of channel enhancement and active sensing, whereas \cite{Arun2020RFocus,ren2022configuring} account for channel enhancement alone.

\begin{table}
[t]\renewcommand\arraystretch{1.5}
\footnotesize
\centering
\caption{List of Main Variables}
\begin{tabular}{|c||l|}
\hline
\textbf{Symbol} & \textbf{Definition} \\ \hline
\hline
$N$ & number of meta-atoms of each MTS \\ \hline
$K$ & number of possible phase shifts per meta-atom \\ \hline
$T$ & number of random samples of RSS\\ \hline
$\omega$ & phase shift spacing $2\pi/K$\\ \hline
$h_{t,r}$ & direct channel from transmitter to receiver \\ \hline
$h_{\ell, u, v}$ & reflected channel of meta-atom $(u,v)$ on MTS $\ell$ \\ \hline
$\overline{h}$ & line-of-sight component of channel $h$ \\ \hline
$\widetilde{h}$ & fading component of channel $h$ \\ \hline
$\theta_{\ell,u,v}$ & phase shift of meta-atom $(u,v)$ of MTS $\ell$ \\ \hline
$\psi_{t,\ell}$ & azimuth angle from the transmitter to MTS $\ell$  \\ \hline
$\psi_{\ell,r}$ & azimuth angle from MTS $\ell$ to the receiver \\ \hline
$\phi_{t, \ell}$ & elevation angle from transmitter to MTS $\ell$  \\ \hline
$\phi_{\ell,r}$ & elevation angle from MTS $\ell$ to receiver\\ \hline
$\mathcal{Q}_{\ell,u,v,k}$ & set of random samples with $\theta_{\ell,u,v}=k\omega$ \\ \hline
$S^{(t)}$ & received signal strength of random sample $t$\\ \hline
$d_M$ & spacing between two adjacent meta-atoms \\ \hline
$(P_x, P_y)$ & position coordinates of transmitter \\ \hline
$(Z_{x,\ell}, Z_{y,\ell})$ & position coordinates of the center of MTS $\ell$ \\ \hline
\end{tabular}
\label{tab:notation}
\end{table}

\emph{Notation:} For a complex number $u$, $\angle u$ refers to the phase of $u$. Let $\mathbb{E}[X]$ be the expectation of the random variable $X$, and let $\widehat{\mathbb{E}}[X]$ be the empirical average. Moreover, write $f(n)=O(g(n))$ if there exists some $c>0$ such that $|f(n)| \leq c g(n)$ for $n$ sufficiently large; write $f(n)=\Omega(g(n))$ if there exists some $c>0$ such that $f(n) \geq c g(n)$ for $n$ sufficiently large; write $f(n)=\Theta(g(n))$ if $f(n)=O(g(n))$ and $f(n)=\Omega(g(n))$ both hold. A summary of variables is provided in Table \ref{tab:notation} for ease of reference.

\section{System Model}
\label{sec:system}

Consider a double-MTS system as shown in Fig.~\ref{fig:system_model}. In principle, each MTS is treated as an anchor to determine one direction of the target, so a pair of MTSs suffices to determine the 2D location, as discussed in greater detail in Section \ref{sec:ISAC}. There is no particular reason for assuming only two MTSs here except the ease of notation; all the results in this paper can be immediately extended to more than two MTSs. Following \cite{mei2021performance,gan2022multiple,yang2022energy}, we neglect those multi-hop cascaded reflected channels for they are much weaker than the direct channel and the single-hop reflected channels. For ease of notation, assume that every MTS consists of $N$ meta-atoms arranged as an $N_{\mathrm{row}}\times N_{\mathrm{col}}$ array, where $N=N_{\mathrm{row}} N_{\mathrm{col}}$. The proposed algorithm can readily account for MTSs with distinct $N$ values, as demonstrated in our field tests shown in Section \ref{sec:experiment}. Meta-atom $(u,v)$ refers to the meta-atom on the $u$th row and the $v$th column of the present MTS, for $u=1,2,\ldots,N_{\mathrm{row}}$ and $v=1,2,\ldots,N_{\mathrm{col}}$. The positions of MTSs are fixed and known \emph{a priori}. The 2D position coordinates of the center of MTS $\ell$ is denoted by $(Z_{x,\ell}, Z_{y,\ell})\in\mathbb R^2$. Moreover, denote by $\theta_{\ell,u,v}$ the phase shift of meta-atom $(u,v)$ of MTS $\ell\in\{1,2\}$. Let $\bm\theta_\ell=\{\theta_{\ell,u,v},\forall u,\forall v\}$ be the set of phase shift variables associated with MTS $\ell$. A $K$-ary discrete constraint is imposed on these phase shift variables:
\begin{equation}
\theta_{\ell,u,v}\in\Phi_K:=\{0,\omega, 2 \omega, \ldots, (K-1) \omega\},
\end{equation}
where $\omega={2 \pi}/{K}$ is the phase shift spacing given a positive integer $K\ge2$. 

\begin{figure}[t]
    \centering
    \includegraphics[width=0.6\linewidth]{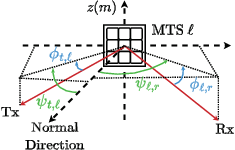}
    \caption{Azimuth and elevation angles from each MTS to transmitter/receiver.}
    \label{fig:model_1}
\end{figure}
\begin{figure}[t]
\centering
\subfigure[]{
\label{fig:model_vertical}
\includegraphics[width=0.4\linewidth]{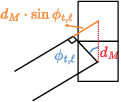}}
\hspace{1em}
\subfigure[]{
\label{fig:model_horizontal}
\includegraphics[width=0.45\linewidth]{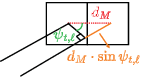}}
\caption{(a) Path difference of incident signal between two vertically adjacent meta-atoms; (b) path difference of incident signal between two horizontally adjacent meta-atoms.}
\end{figure}

Moreover, denote by $h_{t,r}\in\mathbb C$ the direct channel from the transmitter to the receiver, $h_{t,\ell,u,v}\in\mathbb C$ the channel from the transmitter to meta-atom $(u,v)$ of MTS $\ell$, $ h_{\ell,r,u,v}\in\mathbb C$ the channel from meta-atom $(u,v)$ of MTS $\ell$ to the receiver. Each channel $h$ of the above types is modeled as
\begin{equation}
h=\sqrt{\gamma}\left(\sqrt{\frac{\delta}{1+\delta}}\overline{h}+\sqrt{\frac{1}{1+\delta}}\widetilde{h}\right), 
\end{equation}
where $0<\gamma<1$ is the attenuation factor, $
\delta>0$ is the Rician factor, $\overline{h}\in\mathbb C$ is the static component satisfying $|\overline{h}|=1$, and $\widetilde{h}$ is the fading component modeled as a random variable drawn i.i.d. from the complex Gaussian distribution $\mathcal{CN}(0,1)$. Note that the parameters $(\gamma,\delta,\overline{h})$ can vary from channel to channel. The reflected channel associated with meta-atom $(u,v)$ of MTS $\ell$ is given by
\begin{align}
    &h_{\ell,u,v} = h_{t, \ell,u,v}\times h_{\ell,r,u,v},\; \ell=1\text{ or }2.
\end{align}
Further, the line-of-sight (LOS) component is modeled as follows. Denote by $\psi_{t,\ell}$, $\phi_{t,\ell}$, $\psi_{\ell,r}$, and $\phi_{\ell,r}$ the azimuth and elevation angles associated with MTS $\ell$ as shown in Fig.~\ref{fig:model_1}, denote by $\lambda_c$ the carrier frequency, and denote by $d_{M}$ the distance between two adjacent meta-atoms. After a bit of geometry shown in Fig.~\ref{fig:model_vertical} and Fig.~\ref{fig:model_horizontal}, we can obtain the phase of each LOS component as
\begin{align}
    \angle\overline{h}_{t,\ell,u,v} &= \xi d_M\big(v\sin(\psi_{t,\ell})\cos(\phi_{t,\ell})-u\sin(\phi_{t,\ell})\big)\notag +\xi d_{t,\ell}, \notag\\
    \angle\overline{h}_{\ell,r,u,v} &= \xi d_M\big(u\sin(\phi_{\ell,r})-v\sin(\psi_{\ell,r})\cos(\phi_{\ell,r})\big)-\xi d_{\ell,r},\notag\\
    \angle\overline{h}_{t,r} &= \xi d_{t,r},
    \label{eq:LoS_channel}
\end{align}
with the shorthand $\xi = -\frac{2\pi}{\lambda_c}$, where $d_M$ is the spacing between two adjacent meta-atoms, $d_{t,\ell}$ is the distance from the transmitter to MTS $\ell$, $d_{\ell,r}$ is the distance from MTS $\ell$ to the receiver, and $d_{t,r}$ is the distance from the transmitter to the receiver.

As clarified in Section \ref{sec:intro}, we pursue a dual-functional objective of channel enhancement and active sensing. For channel enhancement, we seek the optimal configuration $(\bm\theta_1,\bm\theta_2)$ of the two MTSs to maximize the SNR:
\begin{equation}
\mathrm{SNR}(\bm\theta_1,\bm\theta_2) =\mathbb{E}\Bigg[\Bigg|h_{t,r}+\sum^2_{\ell=1}\sum_{u=1}^{N_{\mathrm{row}}}\sum_{v=1}^{N_{\mathrm{col}}}h_{\ell,u,v}e^{j\theta_{\ell,u,v}}\Bigg|^2\Bigg],
\label{eq:channel_power}
\end{equation}
where the expectation is taken over the fading components $\widetilde{h}$ from all the channels, and the background noise power is normalized to $1$ to ease notation. For active sensing, the receiver aims to acquire the position coordinates $(P_x, P_y)$ of the transmitter based on its RSS.

We reiterate that the CSI, i.e., $\{h_{t,r}, h_{\ell,u,v}\}$, is not available in our problem case, so the SNR function in \eqref{eq:channel_power} is unknown. What we can do at most is to measure the function output $\mathrm{SNR}(\bm\theta_1,\bm\theta_2)$ for each particular choice of $(\bm\theta_1,\bm\theta_2)$, namely the \emph{zero-order optimization} \cite{liu2020primer}.

\section{Channel Enhancement}

Without knowing CSI, a naive method for maximizing the SNR is to try out a sequence of $T$ possible configurations, say $(\bm\theta^{(1)}_1,\bm\theta^{(1)}_2),(\bm\theta^{(2)}_1,\bm\theta^{(2)}_2),\ldots,(\bm\theta^{(T)}_1,\bm\theta^{(T)}_2)$, and in the meanwhile measure their corresponding SNRs---which boil down to the RSS samples $(S^{(1)},S^{(2)},\ldots,S^{(T)})$, and then choose the best $(\bm\theta^{(t)}_1,\bm\theta^{(t)}_2)$ with the highest RSS. However, this greedy search is of extremely low efficiency since it barely utilizes the problem structure. Plus, the greedy search cannot lead us to active sensing.

We now propose a blind configuration method that uses the RSS samples much more efficiently. First, let us group the RSS samples according to $(\bm\theta^{(t)}_1, \bm\theta^{(t)}_2)$: add to the subset $\mathcal{Q}_{\ell,u,v,k}$ all those RSS samples whose phase shift for meta-atom $(u,v)$ of MTS $\ell$ equals $k\omega$, for $k=1,2,\ldots,K$, i.e., 
\begin{equation}
    \mathcal{Q}_{\ell,u,v,k} =\left\{t \mid \theta_{\ell,u,v}^{(t)}=k\omega\right\}.
\end{equation}
We then calculate the \emph{conditional sample mean} of the RSS within each $\mathcal{Q}_{\ell,u,v,k}$:
\begin{equation}
\widehat{\mathbb{E}}\left[S \mid \theta_{\ell,u,v}=k \omega\right]=\frac{1}{\left|\mathcal{Q}_{\ell,u,v,k}\right|} \sum_{t \in \mathcal{Q}_{\ell,u,v,k}}S^{(t)}.
\label{eq:conditional_sample_mean}
\end{equation}
Intuitively speaking, $\widehat{\mathbb{E}}\left[S \mid \theta_{\ell,u,v}=k \omega\right]$ reflects the average goodness of letting $\theta_{\ell,u,v}=k \omega$ when the rest phase shifts are randomly chosen. Thus, we decide each phase shift according to the conditional sample mean:
\begin{equation}
\theta_{\ell,u,v}^{\mathrm{BCM}}=\arg \max _{\varphi \in \Phi_K} \widehat{\mathbb{E}}\left[S \mid \theta_{\ell,u,v}=\varphi\right].
\label{eq:theta_selection_BPSO}
\end{equation}
The above method runs very fast, since both the conditional averaging in \eqref{eq:conditional_sample_mean} and the linear search in \eqref{eq:theta_selection_BPSO} can be finished in $O(KN)$ time. The following theorem shows that the above method guarantees performance for the number of samples $T$ sufficiently large.

\begin{theorem}
\label{thm:SNR}
    When $T=\Omega(N^3 \log N)$, the solution $(\bm\theta_1,\bm\theta_2)$ produced by \eqref{eq:theta_selection_BPSO} leads to $\mathrm{SNR}(\bm\theta_1,\bm\theta_2) = \Theta(N^2)$.
\end{theorem}

The above theorem shows that the solution produced by the proposed algorithm guarantees quadratic scaling of the SNR with respect to the number of meta-atoms. This communication-side analysis extends existing results on single-MTS blind beamforming to the double-MTS setting. Since we only consider the single-hop reflected channels, the two MTSs are equivalent to a single MTS for the SNR boost analysis. The above theorem can follow from Proposition 1 in 
\cite{lai2025adaptive} directly. We sketch the proof here for completeness. First, based on the concentration inequalities including Hoeffding’s inequality and Chebyshev's inequality, we show that each reflected channel can be rotated to the closest possible position to the direct channel by using the phase shift in \eqref{eq:theta_selection_BPSO} provided that $T=\Omega(N^3 \log N)$. Now, with $K\ge2$, we must have the angle between the direct channel and any phase-shifted reflected channel be smaller than or equal to $\pi/2$. Intuitively, all the channels, either direct or reflected, are clustered together, so the total strength grows quadratically with $N$. Note that the above sampling cost $T=\Omega(N^3\log N)$ is higher than the sampling cost $T=\Omega(N^2(\log N)^3)$ in \cite{ren2022configuring} for the static channel case without fading. While Theorem 1 requires a large sample size (albeit polynomial in $N$) to establish a rigorous theoretical bound, $T$ can be significantly smaller in practice. For instance, in our field tests with $N=294$, a much smaller size of $T=3000$ already achieves excellent performance, as shown in Section \ref{sec:experiment}.

\section{Connection to Phase Retrieval}
\label{sec:phase}

In this section, we examine the blind configuration method from a phase retrieval point of view. Simply put, we show that the solution in \eqref{eq:theta_selection_BPSO} is also the optimal solution to the phase retrieval problem with a specific distortion metric. This connection is of practical significance because it reveals that blind configuration is in essence to recover the phase information---which is intimately related to the geographic geometry and thus enables localization.

\begin{figure}[t]
    \centering
    \includegraphics[width=0.8\linewidth]{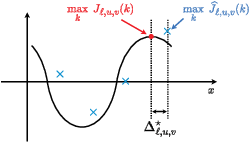}
    \caption{The phase difference $\Delta_{\ell,u,v}$ can be recovered by shifting the trigonometric curve properly to minimize the gap between $\max_{k}J_{\ell,u,v}(k)$ and $\max_{k}\widehat{J}_{\ell,u,v}(k)$.}
    \label{fig:est_Delta}
\end{figure}

We start by introducing a function of index $k=1,\ldots,K$:
\begin{align}
J_{\ell,u,v}(k) &= \mathbb{E}\left[S\mid\theta_{\ell,u,v}=k\omega\right]-\mathbb{E}\left[S\right] \notag \\
&=CA_{\ell,u,v}\cos(k\omega-\Delta_{\ell,u,v}),
\label{real J}
\end{align}
where $C$ is a constant depending on the transmit power, and
\begin{align}
&\Delta_{\ell,u,v} = \angle \overline{h}_{t,r} - \angle\overline{h}_{t,\ell,u,v}-\angle\overline{h}_{\ell,r,u,v}, \label{eq:phase_diff}\\
&A_{\ell,u,v} = 2\sqrt{\frac{\gamma_{t,r}\gamma_{t,\ell,u,v}\gamma_{\ell,r,u,v}\delta_{t, r} \delta_{t,\ell,u,v} \delta_{\ell,r,u,v}}{(1+\delta_{t, r})(1+\delta_{t,\ell,u,v})(1+\delta_{\ell,r,u,v})}}.
\end{align}
Generally speaking, the phase difference $\Delta_{\ell,u,v}$ reflects the AoA with respect to each MTS, while $A_{\ell,u,v}$ reflects the distances between transmitter, MTS, and receiver. In particular, $J_{\ell,u,v}(k)$ quantifies the performance gain of letting $\theta_{\ell,u,v}=k\omega$ over the random choice of $\theta_{\ell,u,v}$, so it is referred to as the \emph{gain function}. Without CSI, the parameters $A_{\ell,u,v}$ and $\Delta_{\ell,u,v}$ in the gain function are unknown.

Note that we can empirically evaluate $J_{\ell,u,v}(k)$ based on the RSS samples as 
\begin{align}
\widehat J_{\ell,u,v}(k) &= \widehat{\mathbb{E}}\left[S\mid\theta_{\ell,u,v}=k\omega\right]-\widehat{\mathbb{E}}\left[S\right] \notag \\
&=\frac{1}{\left|\mathcal{Q}_{\ell,u,v, k}\right|} \sum_{t \in \mathcal{Q}_{\ell,u,v,k}}S^{(t)}-\frac{1}{T} \sum_{t=1}^T S^{(t)}.
\label{eq:compute_Jluvk}
\end{align}
Now we can recover the phase difference $\Delta_{\ell,u,v}$ by comparing $J_{\ell,u,v}(k)$ and $\widehat J_{\ell,u,v}(k)$. As depicted in Fig.~\ref{fig:est_Delta}, with $k$ relaxed to be a continuous variable $x\in\mathbb R$, the gain function $J_{\ell,u,v}(x)$ has a trigonometric curve, while the empirical evaluations $\widehat J_{\ell,u,v}(k)$ correspond to a sequence of discrete sample points at $k=1,2,\ldots,K$. In the ideal case, the peak on the trigonometric curve, $\max_k\;J_{\ell,u,v}(k)$, should coincide with the maximum sample point, $\max_k\;\widehat J_{\ell,u,v}(k)$; in the current case, we shift the trigonometric curve properly by tuning $\Delta_{\ell,u,v}$, so as to minimize the total distortion between $\max_k\;J_{\ell,u,v}(k)$ and  $\max_k\widehat J_{\ell,u,v}(k)$ for all $(\ell,u,v)$, i.e.,
\begin{subequations}
\label{phase_est}
\begin{align}
    \underset{\{\Delta_{\ell,u,v}\}}{\text{minimize}}& \quad \sum_{\ell,u,v}\left|\max_{k}\;J_{\ell,u,v}(k)-\max_{k}\;\widehat{J}_{\ell,u,v}(k)\right|^2\\
    \text {subject to}& \quad\, \Delta_{\ell,u,v}\in(0,2\pi],\;\forall (\ell,u,v).
\end{align}
\end{subequations}
The above problem can be optimally solved as
\begin{equation}
    \label{eq:opt_est_Delta}
    {\Delta}^\star_{\ell,u,v} = k^\star_{\ell,u,v}\omega,
\end{equation}
where
\begin{equation}
\label{opt k}
k^\star_{\ell,u,v}=\arg\max_{k}\;\widehat{J}_{\ell,u,v}(k).
\end{equation}
Furthermore, because $\widehat{\mathbb{E}}\left[S\right]$ is invariant in $k$, we can rewrite \eqref{opt k} as
\begin{equation}
\label{opt_k}
k^\star_{\ell,u,v}=\arg\max_{k}\;\widehat{\mathbb{E}}\left[S\mid\theta_{\ell,u,v}=k\omega\right].
\end{equation}
It is now easy to see that there is a correspondence between the phase retrieval solution \eqref{opt_k} and the blind configuration solution in \eqref{eq:theta_selection_BPSO}. In other words, after computing $\theta_{\ell,u,v}^{\mathrm{BCM}}$, we can immediately obtain ${\Delta}^\star_{\ell,u,v}$. Furthermore, with ${\Delta}^\star_{\ell,u,v}$ in hand, we can recover the position of transmitter by treating the MTSs as anchors, as discussed in the next section.

\begin{figure*}[b]
\hrule
\begin{equation}
    \label{eq:delta_n_expression}
    \Delta_{\ell,u,v}= \xi d_{t,r}-\xi d_M  v\big(\sin(\psi_{\ell,r})\cos(\phi_{\ell,r})-\sin(\psi_{t,\ell})\cos(\phi_{t,\ell})\big) -\xi d_M  u\big(\sin(\phi_{t,\ell})-\sin(\phi_{\ell,r})\big) + \xi (d_{t,\ell}-d_{\ell,r})
\end{equation}
\end{figure*}

\begin{figure}[t]
    \centering
\includegraphics[width=0.7\linewidth]{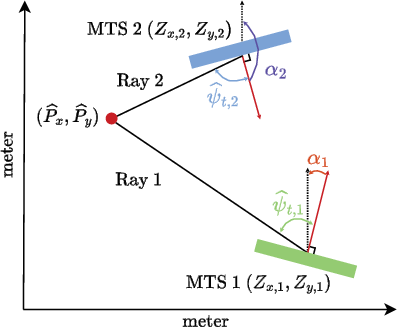}
    \caption{Active sensing by using two MTSs as anchors.}
    \label{fig:est_TX}
\end{figure}

\section{Active Sensing}
\label{sec:ISAC}

\begin{algorithm}[t]
\caption{Blind Configuration Method (BCM)}
\label{alg:active_sensing}
\begin{algorithmic}[1]
    \State \textbf{input:} a series of MTS configuration samples $(\bm\theta_1^{(t)},\bm\theta_2^{(t)})$
    \For{$t=1,2,\ldots,T$}
    \State use the MTS configuration $(\bm\theta_1^{(t)},\bm\theta_2^{(t)})$
    \State measure the RSS $S^{(t)}$
    \EndFor
    \State compute $\widehat{\mathbb{E}}\left[S \mid \theta_{\ell,u,v}=k \omega\right]$ according to \eqref{eq:conditional_sample_mean}
    \State compute $\theta_{\ell,u,v}^{\mathrm{BCM}}$ according to \eqref{eq:theta_selection_BPSO}
    \State recover $\Delta^\star_{\ell,u,v}$ by \eqref{eq:opt_est_Delta}
    \State recover ${\phi}_{t,\ell}$ and ${\psi}_{t,\ell}$ by \eqref{eq:hat_phi} and \eqref{eq:hat_psi}
    \State recover $({P}_x,{P}_y)$ by \eqref{eq:est_Tx_location}.
    \State \textbf{output:} $\theta_{\ell,u,v}^{\mathrm{BCM}}$ and the estimate of $({P}_x,{P}_y)$
\end{algorithmic}
\end{algorithm}

We have shown that the blind configuration method not only boosts the SNR as stated in Theorem \ref{thm:SNR} but also recovers the phase difference $\Delta_{\ell,u,v}$. This section uses the recovered $\Delta_{\ell,u,v}$ to localize the transmitter. First, the phase difference in \eqref{eq:phase_diff} can be rewritten as a function of $({\phi}_{t,\ell},\psi_{t,\ell})$ as shown in \eqref{eq:delta_n_expression} at the bottom of the next page.

Now let us look at two \emph{vertically adjacent} meta-atoms, $(u,v)$ and $(u+1,v)$, as depicted in Fig.~\ref{fig:model_vertical}. With the new expression of $\Delta_{\ell,u,v}$ in \eqref{eq:delta_n_expression}, we can show that any two \emph{vertically adjacent} meta-atoms, $(u,v)$ and $(u+1,v)$, have their phase differences satisfy the following equation:
\begin{equation}
\label{phi difference}
    \Delta_{\ell,u+1,v}-\Delta_{\ell,u,v} =
    \xi d_M\big(\sin(\phi_{\ell,r})-\sin(\phi_{t,\ell})\big).
\end{equation}
Recall that $\phi_{\ell,r}$ depends on the position of MTS relative to receiver, so its value is known to receiver, but the elevation angle $\phi_{t,\ell}$ from MTS to transmitter is unknown. Based on \eqref{phi difference} for the current pair $(u,v)$, we can recover $\phi_{t,\ell}$ as
\begin{equation}
    \widehat\phi_{t,\ell,u,v} = \arcsin\left(\sin (\phi_{\ell,r})-\frac{ \Delta_{\ell,u+1,v}-\Delta_{\ell,u,v}}{\xi d_M}\right).
\end{equation}
The above process shall be repeated for every pair of vertically adjacent meta-atoms, and thus we obtain a sequence of estimates of $\phi_{t,\ell}$ as $\phi_{t,\ell,u,v}$ for the different pairs $(u,v)$. Finally, we average out these estimates to arrive at the ultimate estimate of $\phi_{t,\ell}$ as
\begin{equation}
   \label{eq:hat_phi}
   \widehat{\phi}_{t,\ell} = \frac{1}{(N_{\mathrm{row}}-1)N_{\mathrm{col}}}\sum_{u<N_{\mathrm{row}}}\sum_{v=1}^{N_{\mathrm{col}}}\widehat\phi_{t,\ell,u,v}.
\end{equation}

The azimuth angle $\psi_{t,\ell}$ can be recovered in a similar fashion. Consider two \emph{horizontally adjacent} meta-atoms, $(u,v)$ and $(u,v+1)$, as shown in  Fig.~\ref{fig:model_horizontal}. Observe that their phase differences have this relation:
\begin{multline}
\label{azimuth Delta}
    \Delta_{\ell,u,v+1}-\Delta_{\ell,u,v} = \\
   \xi d_M\big(\sin(\psi_{t,\ell})\cos(\phi_{t,\ell})-\sin(\psi_{\ell,r})\cos(\phi_{\ell,r})\big).
\end{multline}
Both $\psi_{\ell,r}$ and $\phi_{\ell,r}$ are known because they depend on the position of MTS relative to receiver. In light of \eqref{azimuth Delta} for each pair $(u,v)$, we can estimate $\psi_{t,\ell}$ as
\begin{align}
&\widehat\psi_{t,\ell,u,v} = \notag\\
&\;\arcsin\left(\frac{\Delta_{\ell,u, v+1}-\Delta_{\ell,u,v}}{\xi d_M\cos (\widehat{\phi}_{t,\ell})}+\frac{\sin (\psi_{\ell,r})\cos (\phi_{\ell,r})}{\cos (\widehat{\phi}_{t,\ell})}\right).
\end{align}
Again, we average out the above estimate across all possible pairs of horizontally adjacent meta-atoms:
\begin{equation}
   \label{eq:hat_psi}
   \widehat{\psi}_{t,\ell} = \frac{1}{(N_{\mathrm{col}}-1)N_{\mathrm{row}}}\sum\limits_{u=1}^{N_{\mathrm{row}}}\sum\limits_{v<N_{\mathrm{col}}}\widehat\psi_{t,\ell,u,v}.
\end{equation}

With the elevation angle $\phi_{t,\ell}$ and the azimuth angle $\psi_{t,\ell}$ recovered, we can readily localize the transmitter by geometry, as shown in Fig.~\ref{fig:est_TX}. The 2D position coordinates of the transmitter are:
\begin{subequations}
\label{eq:est_Tx_location}
\begin{align}
    &\widehat{P}_x= \notag\\
    &\;\frac{\tan\big(\widehat{\psi}_{t,1}-\alpha_1\big)Z_{x,1}+\tan\big(\widehat{\psi}_{t,2}+\alpha_2\big)Z_{x,2}+Z_{y,1}-Z_{y,2}}{\tan\big(\widehat{\psi}_{t,1}-\alpha_1\big)+\tan\big(\widehat{\psi}_{t,2}+\alpha_2\big)}, \\
    &\begin{aligned}
    \widehat{P}_y &= -\tan\big(\widehat{\psi}_{t,1}-\alpha_1\big)(Z_{x,1} - \widehat{P}_x) +Z_{y,1}, \\
    & =\tan\big(\widehat{\psi}_{t,2}+\alpha_2\big)(Z_{x,2} - \widehat{P}_x)+Z_{y,2}.
    \end{aligned}
\end{align}
\end{subequations}
We now complete the blind configuration method, as summarized in Algorithm \ref{alg:active_sensing}. Regarding the complexity, for an MTS with $N$ meta-atoms, the control overhead required to evaluate $T$ random samples scales linearly with both $N$ and $T$. Meanwhile, the overhead of measuring the received signal power for these $T$ samples at the receiver scales linearly with $T$. Therefore, the complexity of the data acquisition phase is $O(NT)$. Moreover, the complexity after the data acquisition phase is $O(NK)$, so the overall complexity of BCM becomes $O(N(T+K))$.

\begin{remark}
The proposed angle estimation formulas indeed require a LOS component in the transmitter–MTS–receiver path, because the azimuth and elevation angles to be estimated are directly embedded in the phases of the LOS components, as shown in \eqref{eq:LoS_channel}. If the transmitter–MTS–receiver path is strictly dominated by rich diffuse multipath without a clear LOS component, the phase of the reflected channel is no longer directly related to the azimuth and elevation angles, rendering the proposed localization algorithm ineffective. Nonetheless, we wish to remark that this LOS requirement is not specific to our work; it is a standard assumption in most existing studies on MTS-based localization \cite{zhang2021ubiquitous,lin2022localization,keykhosravi2022ris}. Furthermore, this assumption can be easily satisfied in practice through appropriate deployment. For example, in an indoor scenario, the ceiling-mounted MTS proposed in \cite{lai2026followspot} can be utilized to guarantee a clear LOS transmitter–MTS–receiver path. Alternatively, one can create a virtual LOS path by a fractional MTS phase shifting scheme as discussed in \cite{lai2025adaptive}.
\end{remark}

\begin{remark}
Consider a double-MTS-assisted broadcast network in which a single-antenna transmitter sends a common message to $M\geq 2$ single-antenna receivers, while each receiver simultaneously estimates the position of the transmitter. BCM can be extended to this setting as follows. We first run BCM independently for each receiver; the random sampling stage can be performed in parallel across receivers. Receiver $m$ then obtains from BCM an MTS configuration $(\bm{\theta}_1^m,\bm{\theta}_2^m)$ that maximizes its SNR, alongside an estimate of the transmitter position. To enhance the channels for all receivers, we seek a common configuration $(\bm{\theta}_1,\bm{\theta}_2)$ of the two MTSs that maximizes the minimum SNR across receivers. Motivated by this max-min objective, we aggregate the receiver-specific configurations using a majority-voting rule:
\begin{equation}
\theta_{\ell,u,v}
=
\arg\max_{\phi \in \Phi_K}
\sum_{m=1}^{M}
\mathbbm{1}\left\{\theta_{\ell,u,v}^m = \phi\right\}.
\end{equation}
In other words, for each meta-atom, the phase shift that receives the largest number of receiver votes is selected as the final value of $\theta_{\ell,u,v}$. The performance guarantee of this majority-voting strategy has been established in \cite{xu2024blind}.
\end{remark}

\section{Field Tests}
\label{sec:experiment}

\subsection{Testbed and Benchmarks}

\begin{figure}[t]
\centering
\subfigure[MTS model-I]{
\label{fig:MTS_I}
\includegraphics[width=0.45\linewidth]{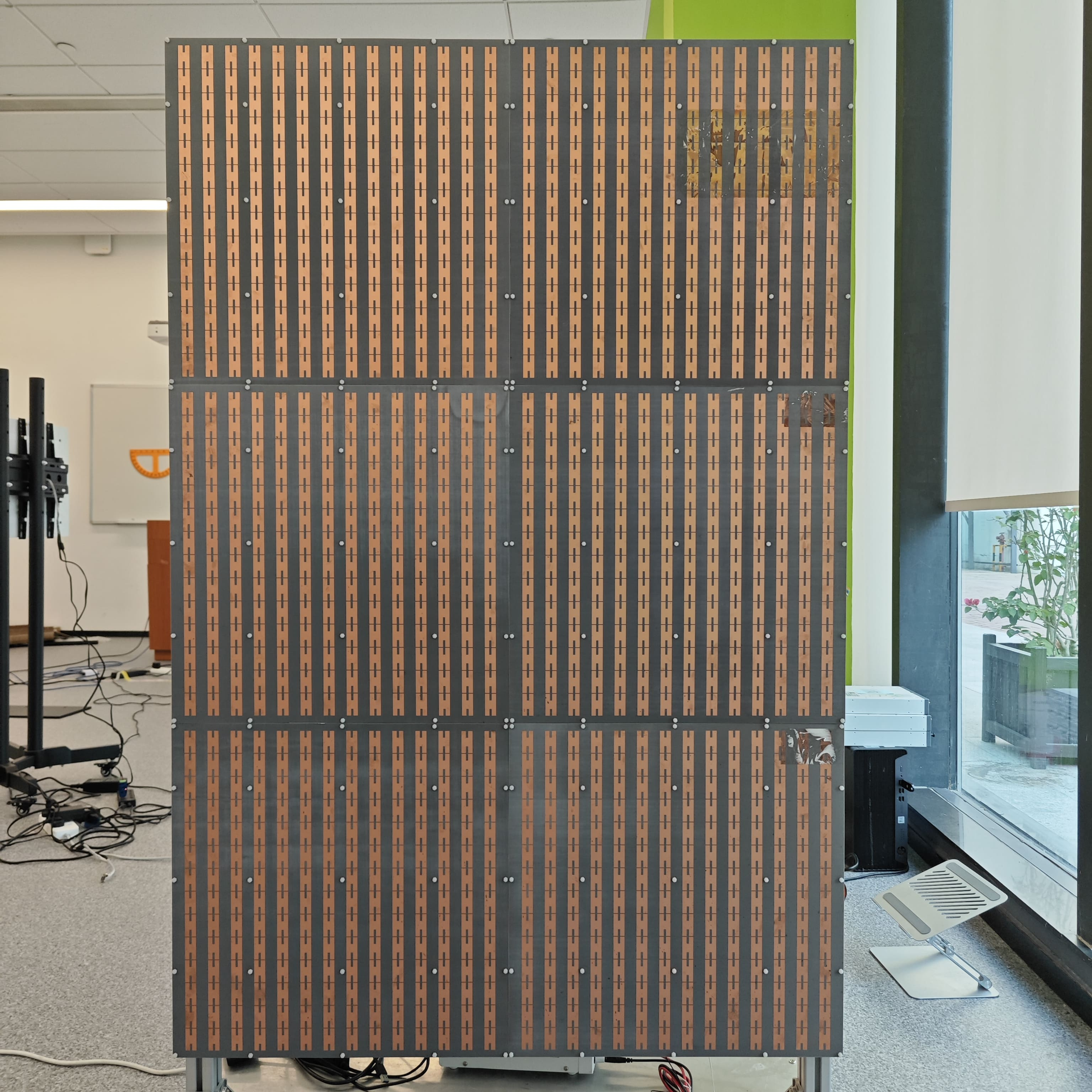}}
\subfigure[MTS model-II]{
\label{fig:MTS_II}
\includegraphics[width=0.45\linewidth]{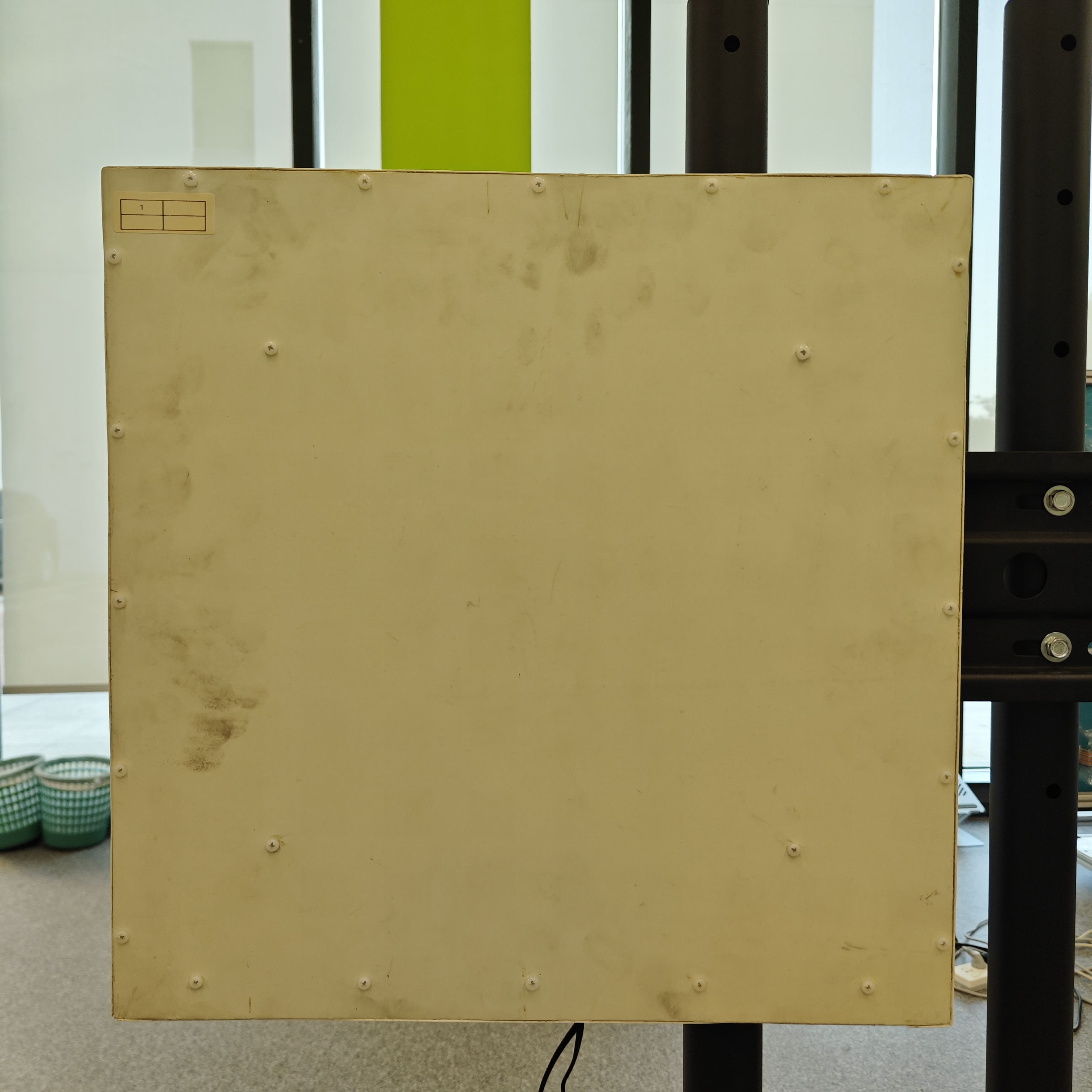}}
\caption{MTS prototype machines used in our field tests.}
\label{fig:MTS_prototype}
\end{figure}

We build a double-MTS system by using a pair of NI-USRP X410 machines as transmitter and receiver, along with the following MTS prototypes as shown in Fig.~\ref{fig:MTS_I} and Fig.~\ref{fig:MTS_II}:
\begin{itemize}
    \item \emph{MTS model-I:} It comprises $21\times 14 = 294$ meta-atoms with spacing of $6$ cm. The phase shift of each meta-atom is chosen from the set $\{0,\pi\}$. Its operating frequency is $2.6$ GHz with a bandwidth of $100$ MHz.
    \item \emph{MTS model-II:} It comprises $10\times 20=200$ meta-atoms with spacing of $4$ cm. The phase shift of each meta-atom is chosen from the set $\{0,{\pi}/{2},\pi, {3\pi}/{2}\}$. Its operating frequency is $3.5$ GHz with a bandwidth of $100$ MHz.
\end{itemize}

We have two MTS prototypes for each model. By default, we consider a system built with MTS model-I. Meter is the unit of distance throughout this section. The transmitter USRP is placed at $(0, 0.53, 0.1)$, the receiver USRP is placed at $(4.51, -0.48, 0.1)$, MTS 1 is placed at $(1.24, -1.25, 1.56)$, and MTS 2 is placed at $(1.24, 1.35, 1.56)$. All the computations are performed on a laptop with one 3.5 GHz AMD Ryzen 7845HS CPU and 32 GB RAM.

The proposed blind configuration method is abbreviated as BCM in the rest of the section. Two separate groups of benchmark methods are considered. For the channel enhancement task, consider these existing methods:
\begin{itemize}
    \item \emph{Zero Phase Shifts (ZPS):} Set every $\theta_{\ell,u,v}$ to zero.
    \item \emph{Beam Scanning \cite{cho2023mmwall,chen2023seamless}:} Try out $T$ random configurations and choose the best.
    \item \emph{DFT:} First estimate the channel by the DFT method in \cite{zheng2019intelligent} and then use the CSI-based global algorithm in \cite{ren2022linear} to optimize phase shifts.
    \item \emph{ALRA:} First estimate the channel by the neural-network-based ALRA method in \cite{yan2024power} and then use the CSI-based algorithm in \cite{ren2022linear} to optimize phase shifts.
\end{itemize}
The performance metric for channel enhancement is:
\begin{equation*}
    \text{SNR Boost} = \frac{\text{SNR with MTSs deployed and configured}}{\text{SNR without MTSs}}. 
\end{equation*}
For the active sensing task, consider the following methods:
\begin{itemize}
    \item \emph{MUSIC:} The receiver estimates the AoA by the MUSIC \cite{stoica1989music}, and estimates the distance based on the channel attenuation, without the MTS deployment.
    
    \item \emph{MUSIC-MTS:} The receiver estimates the AoA by the MUSIC \cite{stoica1989music}, and estimates the distance based on the channel attenuation, with the MTS deployment.
    \item \emph{DFT:} Use the DFT method \cite{zheng2019intelligent} to estimate the phase difference ${\Delta_{\ell,u,v}}$, and then localize the transmitter by using MTSs as anchors.
    \item \emph{ALRA:} Replace the DFT method by the  
    ALRA method \cite{yan2024power} for estimating ${\Delta_{\ell,u,v}}$.
\end{itemize}
The performance metric for active sensing is:
$$
\text{Squared Error} = (P_x-\widehat P_x)^2+(P_y-\widehat P_y)^2.
$$
Moreover, by default, the transmit power is set to $0$ dBm, and the number of samples $T$ is set to $3000$. Regarding the practical time cost, collecting 3000 samples takes 180s in our field tests. We acknowledge that 180 seconds exceeds the \emph{small-scale} channel coherence time. However, this is practically feasible and justified from two perspectives:
\begin{itemize}
    \item \emph{Hardware Prototype Limitation:} The 180s duration is strictly a limitation of our current testbed. State-of-the-art MTS designs (e.g., PIN-diode based) can switch phases in microseconds (e.g., see \cite{li2017holograms}). With mature hardware, acquiring 3000 samples would take merely a fraction of a second, well within the typical channel coherence time.

    \item \emph{Large-Scale Statistical Optimization:} From an algorithmic perspective, our method is robust to long sampling windows. Rather than tracking instantaneous small-scale fading, our algorithm leverages the conditional sample mean to extract slowly varying large-scale channel statistics (e.g., macroscopic spatial geometry and path loss). The extended sampling time naturally averages out the fast-fading fluctuations, yielding highly stable phase configurations that are ideal for robust, long-term channel enhancement and active sensing.
\end{itemize}

We acknowledge that some benchmark methods are not perfectly aligned with our proposed method in terms of the problem setting, primarily because they cannot jointly support channel enhancement and active sensing simultaneously. For example, ZPS and beam scanning are only applicable to the channel enhancement task. Nevertheless, these methods are widely utilized in the literature as standard baselines for their respective individual tasks. Therefore, comparing against them demonstrates that our proposed dual-functional method achieves superior performance even over established, task-specific baselines. Crucially, all methods were evaluated on the exact same hardware prototype system, ensuring a consistent and fair basis for comparison.

\subsection{Influence of Transmit Power}

\begin{figure}[t]
    \centering
    \includegraphics[width=\linewidth]{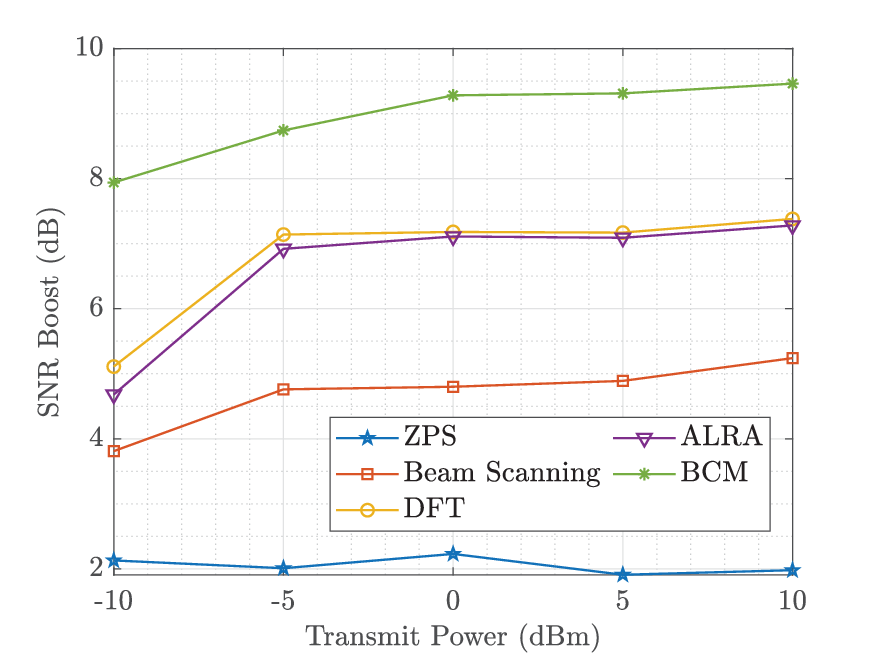}
    \caption{SNR boost vs. transmit power for channel enhancement.}
    \label{fig:snr_vs_p}
\end{figure}

We compare these algorithms when the transmit power level is sequentially set to $-10$ dBm, $-5$ dBm, $0$ dBm, $5$ dBm, and $10$ dBm. Fig.~\ref{fig:snr_vs_p} compares the SNR boost achieved by the different algorithms for channel enhancement. Observe that all the algorithms except ZPS have increasing performance with the transmit power level. Thus, when phase shifts on the MTSs are not adapted to the current wireless environment, the performance gain is always marginal, regardless of the transmit power. The figure also shows that the BCM outperforms the rest algorithms significantly under all the transmit power levels; its advantage is the largest at the low transmit power of $-10$ dBm. The two channel-estimation-based methods, ALRA and DFT, are the second best.

\begin{figure}[t]
    \centering
    \includegraphics[width=\linewidth]{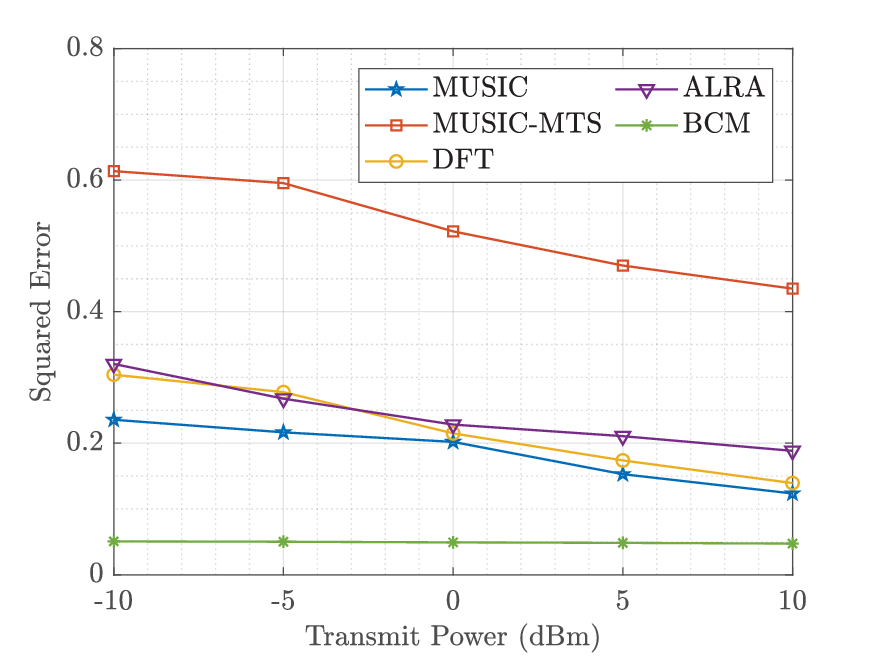}
    \caption{Squared error vs. transmit power for active sensing.}
    \label{fig:error_vs_p}
\end{figure}

We then consider active sensing. Fig.~\ref{fig:error_vs_p} compares the squared error achieved by the different algorithms for active sensing. Observe from the figure that BCM achieves considerably smaller squared error than other algorithms. Again, its advantage over other algorithms is maximal at the lowest transmit power of $-10$ dBm. A surprising fact is that MUSIC-MTS is even worse than MUSIC, so the MTS deployment would degrade MUSIC. The reason is that a large number of extra reflected propagations induced by MTSs can make the wireless environment much more complicated, beyond the capacity of MUSIC.

\begin{figure}[t]
    \centering
    \includegraphics[width=\linewidth]{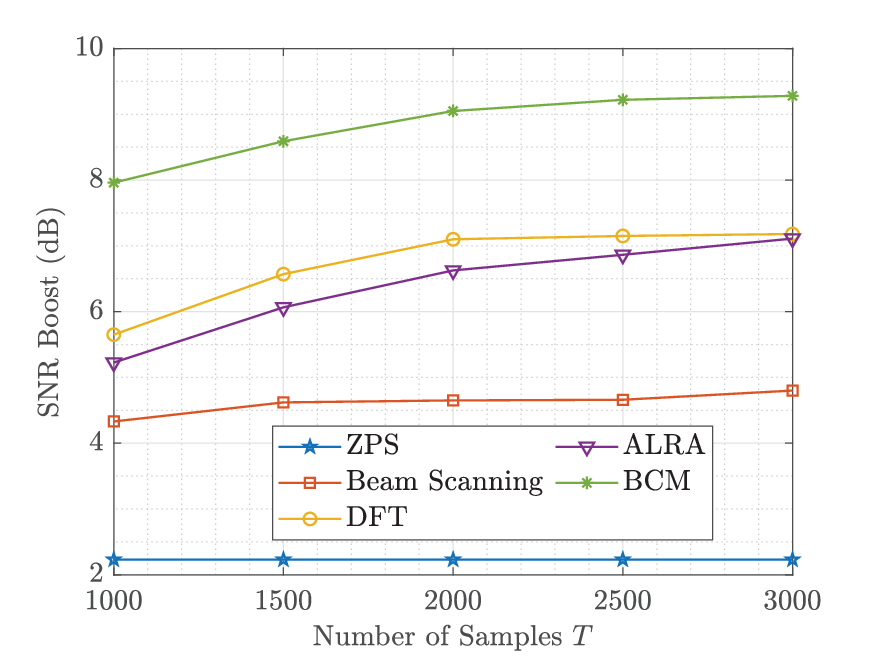}
    \caption{SNR boost vs. $T$ for channel enhancement.}
    \label{fig:snr_vs_T}
\end{figure}

\begin{figure}[t]
    \centering
    \includegraphics[width=\linewidth]{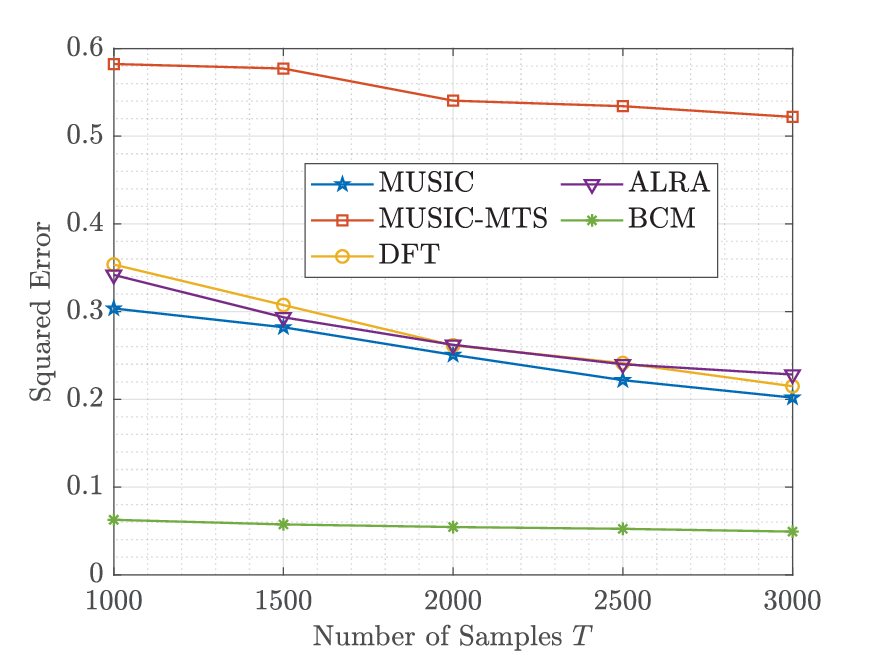}
    \caption{Squared error vs. $T$ for active sensing.}
    \label{fig:error_vs_T}
\end{figure}





\begin{figure*}[t]
\centering
\subfigure[Placement A]{
\begin{minipage}{0.22\linewidth}
    \includegraphics[width=\linewidth]{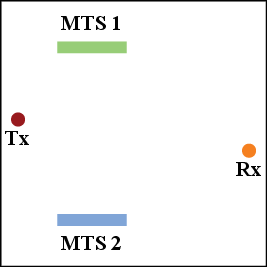}\vspace{1mm}\\
    \includegraphics[width=\linewidth]{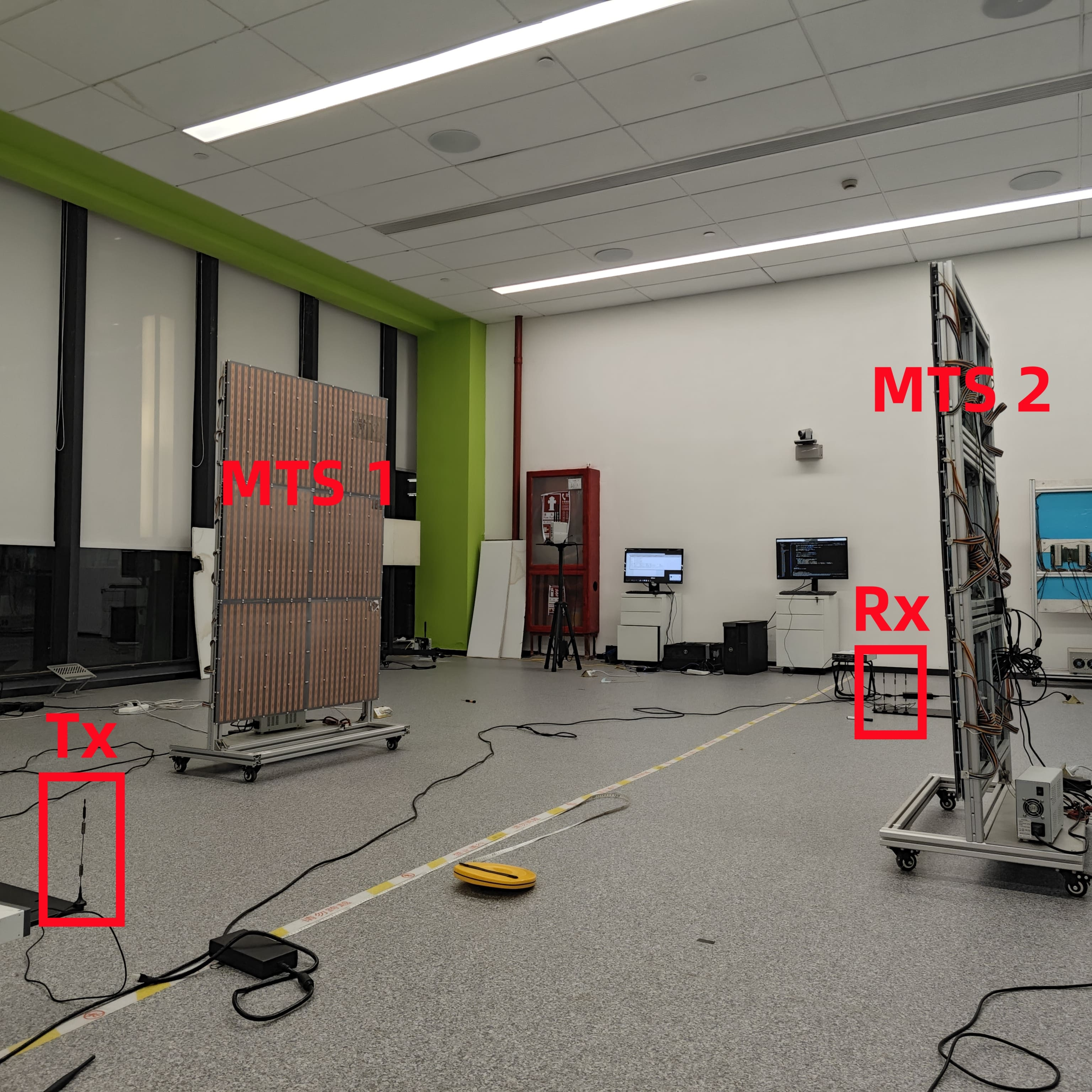}\vspace{1mm}
\end{minipage}
}
\subfigure[Placement B]{
\begin{minipage}{0.22\linewidth}
    \includegraphics[width=\linewidth]{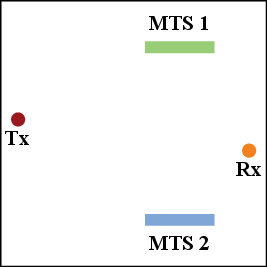}\vspace{1mm}\\
    \includegraphics[width=\linewidth]{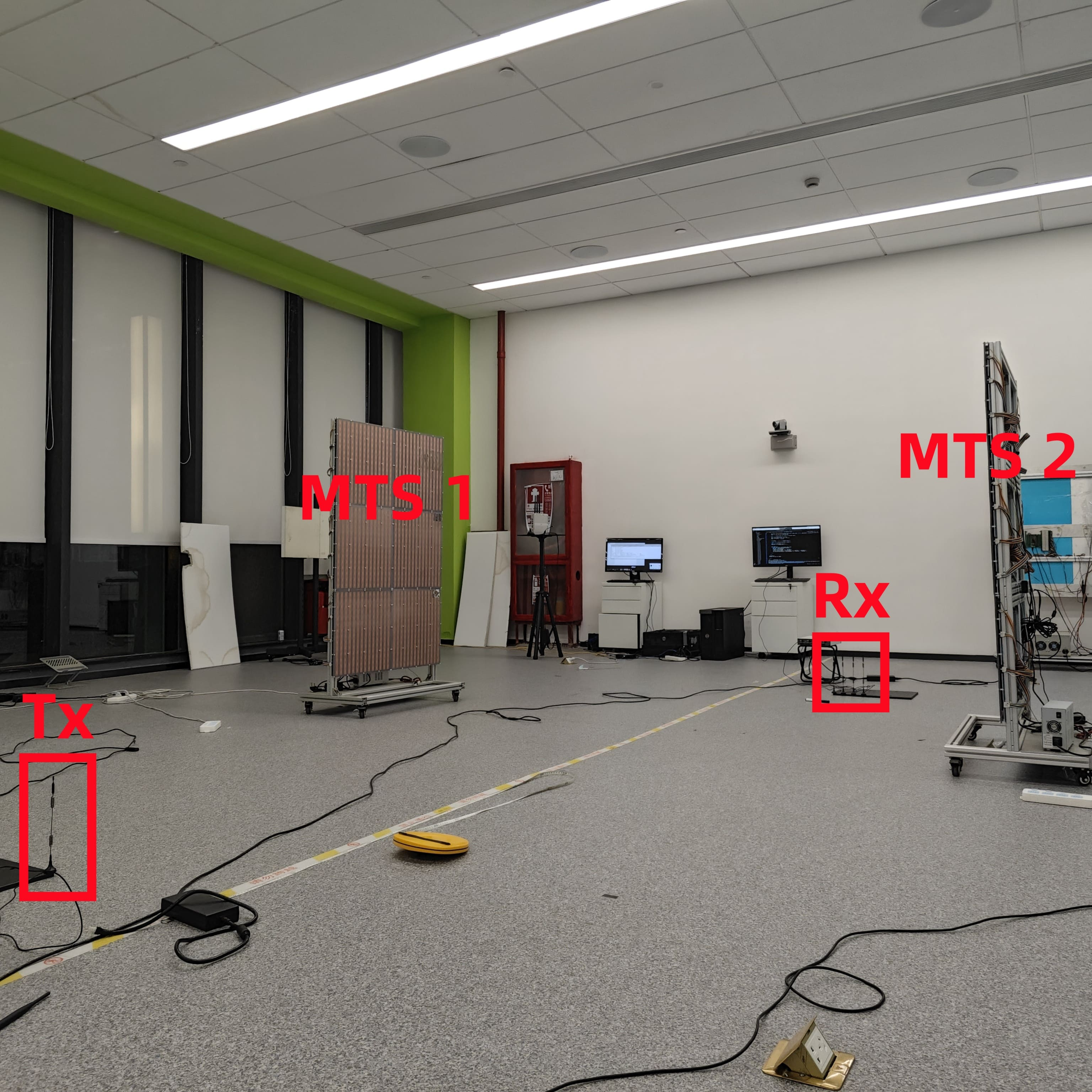}\vspace{1mm}
\end{minipage}}
\subfigure[Placement C]{
\begin{minipage}{0.22\linewidth}
    \includegraphics[width=\linewidth]{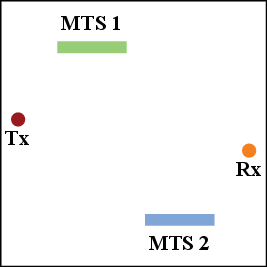}\vspace{1mm}\\
    \includegraphics[width=\linewidth]{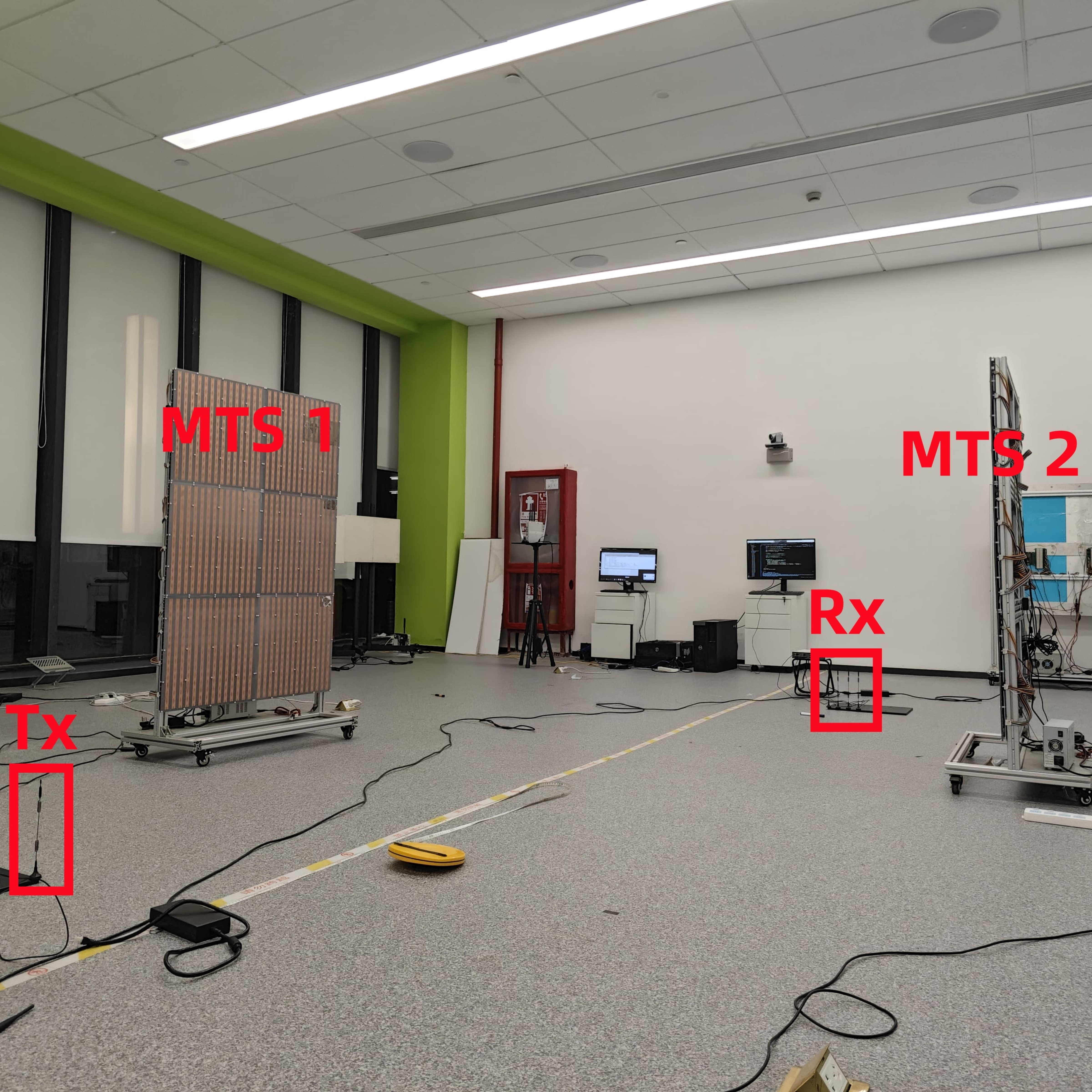}\vspace{1mm}
\end{minipage}}
\subfigure[Placement D]{
\begin{minipage}{0.22\linewidth}
    \includegraphics[width=\linewidth]{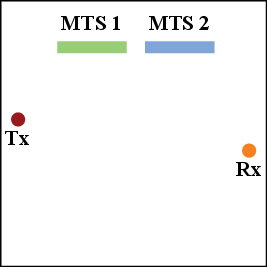}\vspace{1mm}\\
    \includegraphics[width=\linewidth]{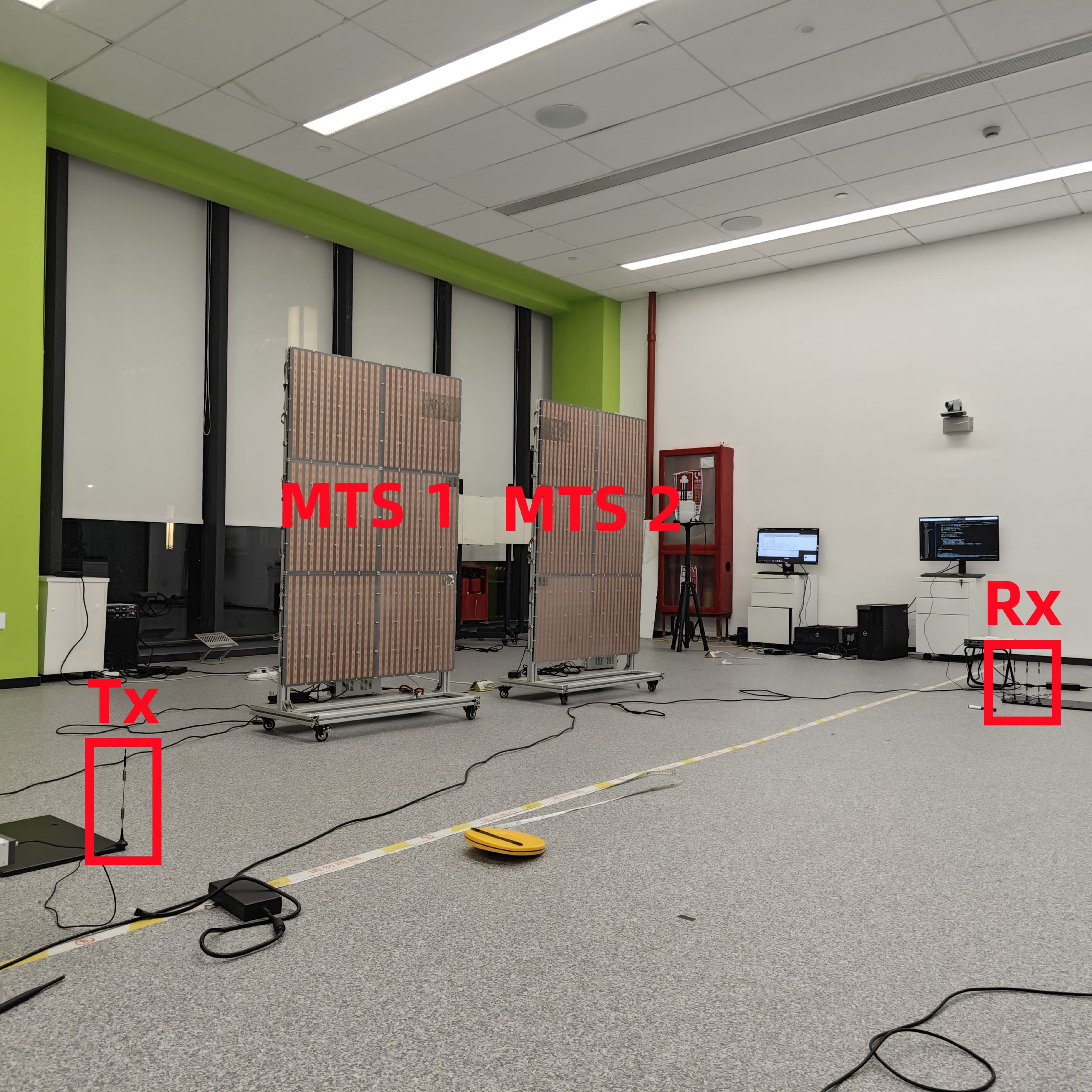}\vspace{1mm}
\end{minipage}}
\caption{Four placements of the two MTSs in our field tests for channel enhancement and active sensing.
}
\label{fig:field_deployment}
\end{figure*}

\begin{table}[t]
\small
    \renewcommand{\arraystretch}{1.3}
\centering
\caption{SNR boost (in dB) for channel enhancement under different MTS parameters
}
\begin{tabular}{lcccc}
\firsthline
& \multicolumn{4}{c}{MTS Parameters $(N, K)$} \\
\cline{2-5}
Method    & $(200, 4)$ & $(200, 2)$ & $(100, 4)$ & $(100, 2)$ \\
\hline
ZPS                & $-1.19$ & $-1.19$ & $-0.08$ & $-0.08$ \\
Beam Scanning                & $4.13$  & $3.77$  & $2.79$  & $2.68$  \\
DFT  & $6.28$  & $4.95$  & $4.76$  & $4.38$  \\
ALRA               & $6.05$  & $4.91$  & $4.42$  & $4.04$  \\
BCM & $8.65$  & $6.50$   & $6.22$  & $4.95$  \\
\lasthline
\end{tabular}
\label{tab:boost_NK}
\end{table}

\begin{table}[t]
\small
\renewcommand{\arraystretch}{1.3}
\centering
\caption{Squared error for active sensing under different\\ MTS parameters
}
\begin{tabular}{lcccc}
\firsthline
& \multicolumn{4}{c}{MTS Parameters $(N, K)$} \\
\cline{2-5}
Method   & $(200, 4)$ & $(200, 2)$ & $(100, 4)$ & $(100, 2)$ \\
\hline
MUSIC  & N/A & N/A & N/A & N/A \\
MUSIC-MTS  & $1.1464$ & $1.1464$ & $0.7980$ & $0.7980$ \\
DFT   & $0.1668$ & $0.2183$&$0.1946$& $0.2453$ \\
ALRA   & $0.1777$ & $0.2294$ &$0.2062$  &$0.2598$  \\
BCM & $0.0260$ & $0.0706$ & $0.0492$ & $0.0988$ \\
\lasthline
\end{tabular}
\label{tab:MSE_NK}
\end{table}

\subsection{Influence of Number of Samples $T$}

In this part, we vary the number of random samples $T$. Recall that $T$ is a design parameter for all the competitor algorithms except ZPS.

Fig.~\ref{fig:snr_vs_T} shows the performance in channel enhancement. As can be expected, increasing $T$ can improve performance so long as the algorithm depends on $T$. The gap between BCM and the second best algorithm DFT does not change much with $T$. In contrast, Beam Scanning has extremely slow improvement as $T$ increases. This fact agrees with the previous claim that optimizing phase shifts in a greedy fashion leads to low efficiency.

Moreover, as shown in Fig.~\ref{fig:error_vs_T}, we can see that BCM continues to outperform the rest algorithms significantly in active sensing. The performance of BCM is fairly stable with respect to $T$. This is because BCM has fast convergence; $1000$ samples are sufficient for BCM to get close to convergence. Again, MUSIC-MTS results in the worst performance.

\subsection{Influence of MTS Parameters}

We now rebuild the system by using MTS model-II. If MTS model-II is fully activated, then there are $N=200$ meta-atoms on each MTS, and each meta-atom has $K=4$ options for phase shift. We can also switch MTS to the half-activation mode so that $N$ reduces to $100$. Besides, we can remove $\pi/2$ and $3\pi/2$ from the phase shift options so that $K$ reduces to 2. As a consequence, we can impose four different parameter settings for $(N,K)$: $(200,4)$, $(200,2)$, $(100,4)$, and $(100,2)$.

We now run the different algorithms for each of the above four parameter settings of MTS. Table \ref{tab:boost_NK} shows the case of channel enhancement. It can be seen that BCM achieves the best performance under all the parameter settings. When the number of meta-atoms is doubled for $K=4$, the performance of BCM is improved by $39$\%; when the number of phase shift options is doubled for $N=200$, the performance of BCM is improved by $33$\%. It is surprising to see that ZPS becomes even worse when MTSs become larger. This is because the introduction of more reflections without proper phase shift optimization needs not bring performance gain. We also consider the case of active sensing in Table \ref{tab:MSE_NK}. Observe that BCM with $(N,K)=(100,2)$ can already outperform the rest algorithms with $(N,K)=(200,4)$. Notice that the performance of MUSIC-MTS is independent of $K$ since it simply fixes every phase shift at zero.

\begin{table}[t]
\centering
\renewcommand\arraystretch{1.2}
\small
\caption{SNR boost (in dB) for channel enhancement under different MTS placements
}
\begin{tabular}{lcccc}
\firsthline
& \multicolumn{4}{c}{MTS Placement} \\
\cline{2-5}
Method   & A & B & C & D \\
\hline
ZPS     & $2.23$ & $-0.93$ & $-0.85$ & $2.13$ \\
Beam Scanning     & $4.80$  & $1.83$  & $1.99$  & $4.30$ \\
DFT     & $7.68$ & $6.18$  & $4.37$  & $6.29$ \\
ALRA    & $7.11$ & $5.56$  & $4.12$  & $6.06$ \\
BCM & $9.28$ & $8.51$  & $7.88$  & $10.01$ \\
\lasthline
\end{tabular}
\label{tab:deployment_boost}
\end{table}

\begin{table}[t]
\renewcommand\arraystretch{1.2}
\small
\centering
\caption{Squared error for active sensing under different\\ MTS placements
}
\begin{tabular}{lcccc}
\firsthline
& \multicolumn{4}{c}{MTS Placement} \\
\cline{2-5}
Method   & A & B & C & D \\
\hline
MUSIC  & N/A & N/A & N/A & N/A \\
MUSIC-MTS         & $0.5220$  & $0.7066$ & $0.6139$ & $0.5929$ \\
DFT                & $0.2149$ & $0.4614$ & $0.2918$ & $0.6947$ \\
ALRA               & $0.2283$ & $0.5135$ & $0.3039$ & $0.7546$ \\
BCM & $0.0492$ & $0.1098$ & $0.0923$  & $0.3854$ \\
\lasthline
\end{tabular}
\label{tab:deployment_MSE}
\end{table}

\subsection{Influence of MTS Placement}

We now consider different placements of the two MTSs as shown in Fig.~\ref{fig:field_deployment}. Table \ref{tab:deployment_boost} shows the channel enhancement performance of the different algorithms for the different MTS placements. While BCM reaches the highest SNR boost under the MTS placement D, the rest algorithms have their best performances under the MTS placement A. For BCM, the gap between the best placement and the worst placement is $2.13$ dB. In contrast, the gap of ZPS is $3.16$ dB, the gap of Beam Scanning is $2.97$ dB, the gap of DFT is $3.31$ dB, and the gap of ALRA is $2.99$ dB. Thus, BCM not only achieves the best performance, but also provides the most stable performance when the MTS placement is altered. A similar conclusion can be obtained from Table \ref{tab:deployment_MSE} for the active sensing task.

\begin{figure}[t]
\centering
\subfigure[LOS case]{
\label{fig:scenario_LoS}
\includegraphics[width=0.46\linewidth]{figs/Deployment_A.eps}}
\subfigure[NLOS case]{
\label{fig:scenario_NLoS}
\includegraphics[width=0.46\linewidth]{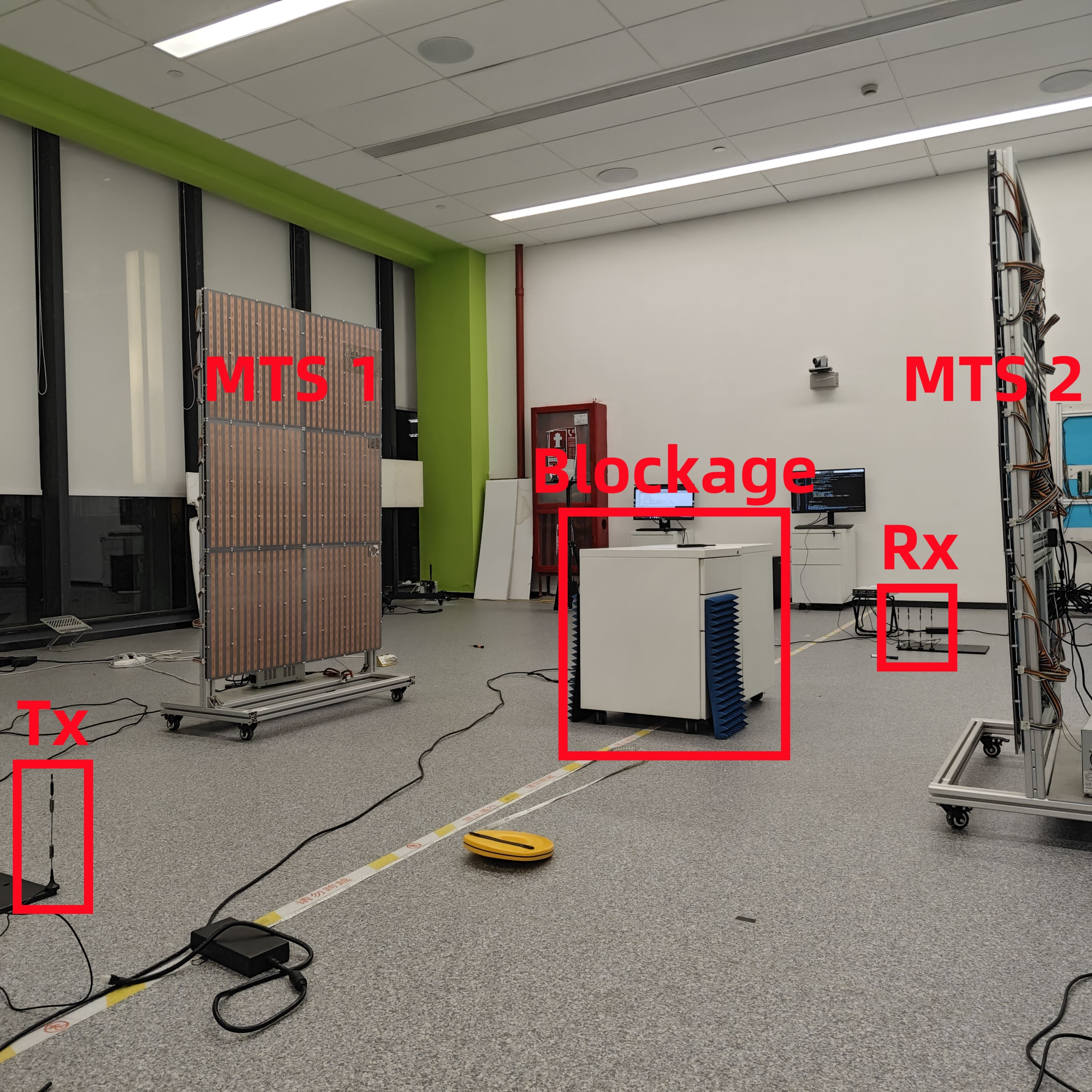}}
\caption{(a) Transmitter and receiver have a LOS direct channel between them; (b) the direct channel is blocked so it becomes NLOS.}
\end{figure}

\subsection{Line-of-Sight vs. Non-Line-of-Sight}

Next, we would like to see how the direct channel impacts the performance. Consider the LOS case shown in Fig.~\ref{fig:scenario_LoS}, and the non-line-of-sight (NLOS) case shown in Fig.~\ref{fig:scenario_NLoS} where the direct propagation from transmitter to receiver is blocked.

\begin{figure}[t]
    \centering
    \includegraphics[width=\linewidth]{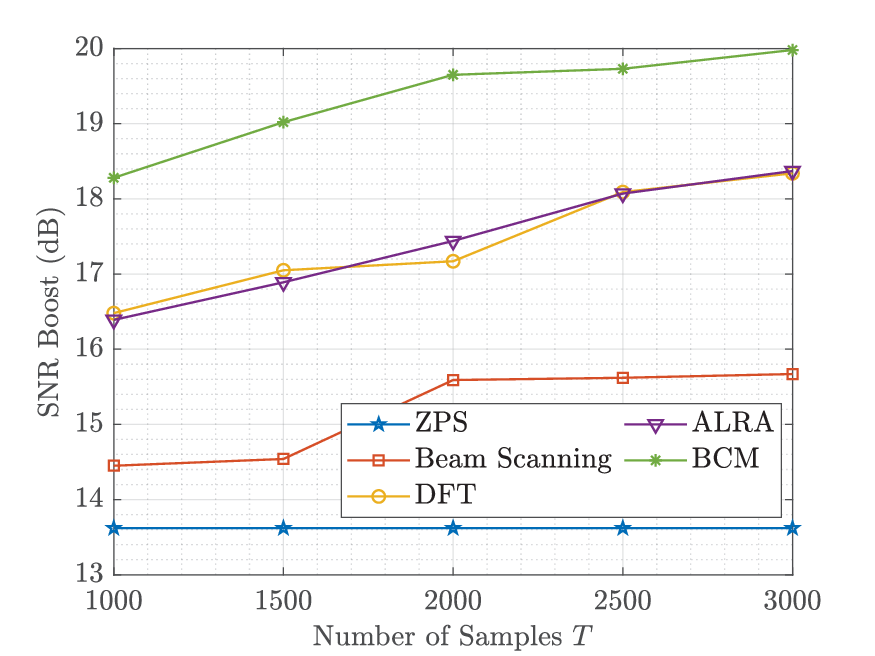}
    \caption{SNR boost vs. $T$ for channel enhancement in the NLOS case.}
    \label{fig:boost_NLoS}
\end{figure}

For the performance in channel enhancement, we compare  Fig.~\ref{fig:snr_vs_T} and Fig.~\ref{fig:boost_NLoS}, with different values of $T$ considered. As can be expected, the performances of all the algorithms considerably drop when the direct channel is blocked. Nevertheless, the trends of most curves are similar to those in the LOS case, only that the curve of Beam Scanning becomes less steep in the NLOS case.

\begin{figure}[t]
    \centering
    \includegraphics[width=\linewidth]{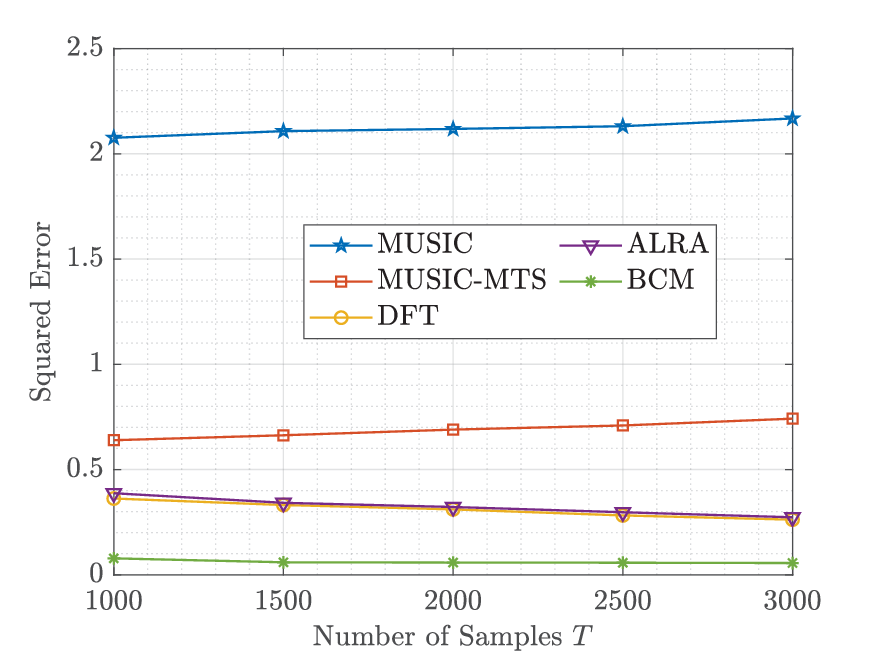}
    \caption{Squared error vs. $T$ for active sensing in the NLOS case.}
    \label{fig:MSE_NLoS}
\end{figure}

\begin{table}[t]
\small
\renewcommand{\arraystretch}{1.3}
\centering
\caption{Run time of different methods for achieving both channel enhancement and active sensing}
\begin{tabular}{lrrrr}
\firsthline
& \multicolumn{3}{c}{Running Time (second)}\\
\cline{2-4}
Method      & $T=1000$ & $T=2000$ & $T=3000$\\
\hline
DFT   & $0.107$  & $0.123$ &  $0.152$\\
ALRA  & $37.71$  & $101.19$ & $490.12$ \\
BCM & $0.048$  & $0.93$ & $0.141$ \\
\lasthline
\end{tabular}
\label{tab:run_time}
\end{table}

In contrast, the trends of curves are quite different between LOS and NLOS when it comes to the active sensing task, as shown in Fig.~\ref{fig:error_vs_T} and Fig.~\ref{fig:MSE_NLoS}. The curves of MUSIC and MUSIC-MTS flatten out in the NLOS case. This fact shows that it is much more difficult to acquire the AoA when the direct channel is too weak. But BCM still leads to excellent performance in the NLOS case. Observe from Fig.~\ref{fig:MSE_NLoS} that its squared error is fairly close to zero.


\subsection{Run Time of Different Algorithms}

We further compare the time efficiencies of the different algorithms. Note that aside from the proposed BCM, only DFT \cite{zheng2019intelligent} and ALRA account for both channel enhancement and active sensing. Clearly, the random sample size $T$ is critical to the running, so we consider running time under different values of $T$ in Table \ref{tab:run_time}. Observe that ALRA requires far more running time than the other methods. DFT and BCM have similar efficiencies overall; the latter runs faster when $T$ is small. Thus, BCM not only achieves significantly better performance (in both tasks) than the existing methods, but also provides high efficiency.

\section{Conclusion}
\label{sec:conclusion}
This work proposes a statistics-based blind configuration method for MTS for achieving channel enhancement and active sensing simultaneously. Differing from the conventional CSI-based methods,  the blind configuration method purely relies on a random set of the RSS samples without any channel knowledge, namely the zero-order optimization. Our method is based on two insights: first, it suffices to consider the conditional sample mean of the RSS in solving the SNR maximization problem; second, the above solution is intimately related to phase retrieval---which is a key enabler for localization. Because the proposed blind configuration method does not require any CSI and only entails lite computation, it is particularly suited for the plug-and-play deployment of MTSs, as demonstrated in the extensive field tests shown in the paper.

\bibliographystyle{IEEEtran}     
\bibliography{IEEEabrv,ref}

\begin{IEEEbiographynophoto}{Wenhai Lai}(Member, IEEE) received the B.E. degree in information engineering from Beijing University of Posts and Telecommunications in 2021 and the Ph.D. degree in computer and information engineering from the School of Science and Engineering, The Chinese University of Hong Kong, Shenzhen, China in 2026. He has been an assistant professor at the School of General Education at Dalian University of Technology since 2026. His research interests include metasurfaces, optimization, and machine learning.
\end{IEEEbiographynophoto}

\begin{IEEEbiographynophoto}{Kaiming Shen} (Senior Member, IEEE) received the B.Eng. degree in information security and the B.Sc. degree in mathematics from Shanghai Jiao Tong University, China in 2011, and then the Ph.D. degree in electrical and computer engineering from the University of Toronto, Canada in 2020. He is a now a tenured Associate Professor with the School of Science and Engineering at The Chinese University of Hong Kong, Shenzhen, China. His research interests include optimization, wireless communications, and information theory. He currently serves as an Editor for IEEE Transactions on Wireless Communications. He is a member of the Signal Processing for Communications and Networking (SPCOM) Technical Committee of the IEEE Signal Processing Society. He received the IEEE Signal Processing Society Young Author Best Paper Award in 2021, the University Teaching Achievement Award in 2023, the Frontiers of Science Award in 2024, and the Chinese Information Theory Society Young Researcher Award in 2025.
\end{IEEEbiographynophoto}

\end{document}